\documentclass[a4paper,10pt,twoside]{cpc-hepnp}

\usepackage{multicol}
\usepackage{graphicx}
\usepackage{booktabs}
\usepackage{amssymb,bm,mathrsfs,bbm,amscd}
\usepackage[tbtags]{amsmath}
\usepackage{lastpage}

\begin{document}

\fancyhead[co]{\footnotesize Author1~ et al: Instruction for typesetting manuscripts}

\footnotetext[0]{Received 14 January 2010}

\title{Status of the Fermilab Muon $(g-2)$ Experiment\thanks{
    FERMILAB-CONF-10-012-E: Supported in
    part by the U.S. National  Science Foundation and 
the U.S. Department of Energy} }

\author{B. Lee Roberts$^{1,2;1)}$\email{roberts@bu.edu}%
}
\maketitle

\address{%
1~(Department of Physics, Boston University, Boston, MA 02215, USA)\\
2~(On behalf of the New Muon $(g-2)$ Collaboration)\\
}

\begin{abstract}
The New Muon $(g-2)$ Collaboration at Fermilab has proposed to measure the
anomalous magnetic moment of the muon, $a_\mu$, a factor of four better than was
done in E821 at the Brookhaven AGS, which obtained 
$a_\mu = [116\,592\,089 (63)] \times 10^{-11}$ $\pm 0.54$~ppm.  The last digit
of $a_{\mu}$ is changed from the published value owing to
a new value of the ratio of the muon-to-proton magnetic
moment that has become available. 
 At present
there appears to be a difference between the Standard-Model value and the
measured value, at the $\simeq 3$  standard deviation level when
electron-positron annihilation data are used to determine the lowest-order
hadronic piece of the Standard Model contribution. The improved
experiment, along with further advances in the determination of the hadronic
contribution, should clarify this difference.  Because of its ability to
constrain the interpretation of discoveries made at the LHC, the improved
measurement will be of significant value, whatever
discoveries may come from the LHC. 
\end{abstract}



\begin{multicols}{2}

\section{Introduction}

The anomalous magnetic moment (anomaly)
 of the $e$, $\mu$  or $\tau$ lepton 
is defined by 
\begin{equation}
a_{e,\mu,\tau} = \frac {(g_{e,\mu,\tau}-2)}{2}\, ;  \ \ 
 \vec{\mu}_{e,\mu,\tau}=g_{e,\mu,\tau}\left( \frac{Qe}{2m_{e,\mu,\tau}}\right) \vec{s}\, ,
\end{equation} 
where $\vec \mu$ is the magnetic dipole moment, and the factor $g$ is
equal to $2$ in the Dirac theory. One of the important discoveries on the
path to the development of QED, and then the Standard Model, was the
measurement by Kusch and Foley~\cite{blr-Kusch}  which showed
definitively that $g_e > 2$. Almost simultaneously,  Schwinger showed that
this difference could be explained by the (one-loop in modern language)
radiative correction with the value $\alpha/2 \pi \simeq 0.00116\cdots$,
independent of the lepton mass.  

The Standard-Model value of $a_\mu$
 arises from loop contributions containing virtual
 photons, leptons, gauge bosons, and hadrons in vacuum polarization loops.
 Other talks at this
meeting have discussed the Standard-Model contributions in some detail.
 For a
general review the reader is referred to the review article by Miller, 
et al.,~\cite{blr-MdRR2007}.  

The muon anomaly has been measured in a series of experiments that began over
fifty years ago\cite{blr-Garwin}, the most recent, E821 at the Brookhaven
AGS,  achieving a precision of 
of $\pm 0.54$ parts per million (ppm)
~\cite{blr-bennett04,blr-bennett06}: 
\begin{equation}
a_\mu = [116\,592\,089 (63)] \times 10^{-11}\, 0.54\,{\rm ppm}.
\label{blr-eq:amuE821}
\end{equation}
The result has been slightly adjusted from the value reported in
Ref.~\cite{blr-bennett04,blr-bennett06} because the value of 
the fundamental constant
$\lambda = \mu_\mu/\mu_p$, the muon to proton magnetic moment ratio,
(see Eq.~(\ref{blr-eq:lambda})),
has changed~\cite{blr-CODATA08}.  The statistical error in the anomaly 
 is $\pm 0.46$~ppm and the systematic error is
$\pm 0.28$~ppm.  The goal of the new Fermilab experiment~\cite{fermilabP989}
 is equal statistical
and systematic errors of $\pm 0.1$~ppm, for a combined error of 0.14~ppm.

Interestingly enough, the measured
 muon anomaly seems to be slightly larger than the
Standard-Model value of~\cite{blr-Davier09} 
\begin{equation}
a_\mu^{\rm SM}[e^+e^-] = 116\, 591\, 834 (49) ]\times 10^{-11}
\label{blr-eq:amuSM}
\end{equation}
which uses $e^+e^-$ annihilation into hadrons to determine the hadronic
contribution, and the value of Prades et al.,~\cite{blr-PdRV} for the hadronic
light-by-light contribution. There is  a difference of $\sim3.2~\sigma$
between the two.  If hadronic $\tau$ decays are used
to determine the lowest-order hadronic contribution (a determination that
relies on significant isospin corrections), the difference drops to
$\sim 2 \sigma$~\cite{blr-Davier09-tau}.

Non-Standard-Model contributions could come from muon substructure,
supersymmetry or extra dimensions, to name a few possibilities. 
Excellent reviews on this topic 
have been written by St\"ockinger~\cite{blr-stockinger}, and
Czarnecki and Marciano~\cite{blr-czmar}.  The SUSY contribution depends on 
$\tan \beta$ and the sign of the
$\mu$ parameter~\cite{blr-stockinger,blr-czmar}: 
\begin{equation}
a_\mu^{\rm SUSY}
 \simeq \, {\rm sgn}\, \mu \, 130\times 10^{-11} \left( 100 {\rm\ GeV}\over 
\widetilde{m}\right)^2 \tan\beta .
\end{equation}

Both $\tan \beta$ and the $\mu$ parameter will be difficult to determine at
LHC. The sign of the deviation of $a_\mu$ from the Standard Model gives the
sign of $\mu$,  and the plot below illustrates the sensitivity of LHC and
$a_\mu$ to $\tan \beta$.  It assumes that the 
SPS1a scenario is realized at LHC~\cite{blr-stockinger}.
The difference between the Standard Model and the result from E821
 is assumed to be 
$\Delta_{a_\mu}= (255 \pm 80) \times 10^{-11}$.  The band labeled ``Fermilab'' 
assumes the same $\Delta_{a_\mu}$ but with an error of 
$\pm 34 \times 10^{-11}$  The improved error comes from the projected 0.14 ppm experimental
error, and improved knowledge of the hadronic contribution to $a_\mu$.
 See Ref.~\cite{blr-stockinger,blr-whitepaper} for more details.

\begin{center}
\includegraphics[width=7.5cm]{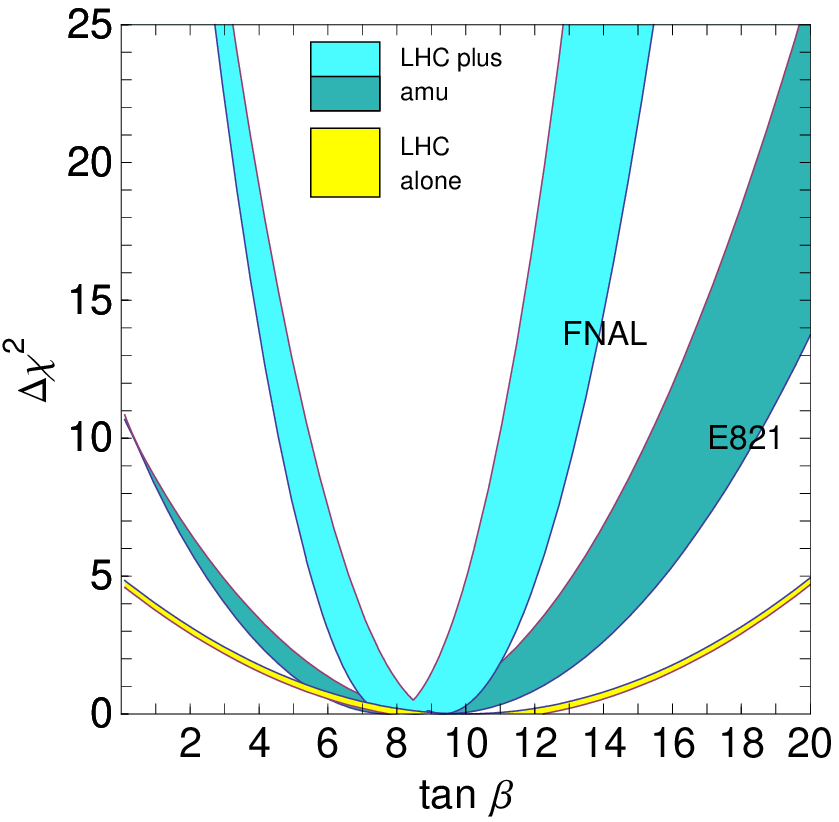}
\figcaption{\label{fg:blr-blueband} A ``blueband'' plot showing the LHC and 
muon $(g-2)$ sensitivities to $\tan \beta$. The   (Figure courtesy of
D. St\"ockinger)}
\end{center}

\section{Measuring $a_\mu$}

The measurement of $a_\mu$ uses the spin precession resulting from the torque
experienced by the magnetic moment when placed in a magnetic field. An
ensemble
 of polarized
muons is introduced into a magnetic field, where they are stored for the
measurement period.  Assuming
that the muon velocity is transverse to the magnetic field ($\vec \beta \cdot
\vec B=0$), the rate at which
the spin turns relative to the momentum vector is given by the difference
frequency between the spin precession and cyclotron frequencies.  With an
electric field present as well as a magnetic one, the difference frequency
becomes
\begin{eqnarray}
\vec \omega_a &=& \vec \omega_S - \vec \omega_C \nonumber \\
&=& - \ \frac{Qe }{ m} 
\left[ a_{\mu} \vec B -
\left( a_{\mu}- \frac{1 }{ \gamma^2 - 1} \right)
\frac{ {\vec \beta \times \vec E }}{ c }
\right]\, ,
\label{eq:blr-omegaa}
\end{eqnarray}
 where $\gamma = (1 -\beta^2)^{-\frac{1}{2}}$.
(The reason for introducing an electric field will become apparent in the
next section.)
The experimentally measured numbers are the muon spin frequency $\omega_a$
and the magnetic field, which is measured with proton NMR, calibrated to the
Larmor precession frequency, $\omega_p$, of a {\it free} proton.
The anomaly is related to these two frequencies by
\begin{equation}
a_{\mu} 
=  \frac {\tilde \omega_a / \omega_p} 
   { \lambda -  {\tilde \omega_a / \omega_p}}
=
\frac {\mathcal R} { \lambda -  {\mathcal R}}  ,
\label{blr-eq:lambda}
\end{equation}
 $\lambda = \mu_{\mu} / \mu_p = 3.183\, 345\,137(85)$, and
${\mathcal R}= \tilde \omega_a / \omega_p$. The tilde over
$\omega_a$ means it has been corrected for the electric-field and pitch
($\vec \beta \cdot \vec B \neq 0$)
corrections~\cite{blr-MdRR2007}.  The ratio $\lambda$ is
determined experimentally from the hyperfine structure of muonium,
the $\mu^+e^-$ atom~\cite{blr-liu,blr-CODATA08}. As mentioned
above, the recommended value
of $\lambda$ has changed slightly since the final results of E821 were
published~\cite{blr-bennett04,blr-bennett06}, 
increasing the value of $a_\mu $ by $9 \times 10^{-11}$,
which is reflected in Eq.~(\ref{blr-eq:amuE821}).

\subsection{The Magic-$\gamma$ Technique}

In the 2001 data set, the systematic errors on the magnetic field
were reduced to $0.17$~ppm.  A number of contributions
 went into this small error,
but one which we wish to emphasize here is the average magnetic field
experienced by the muon ensemble. The magnetic field in
Eq.~(\ref{eq:blr-omegaa}) is an average that can be expressed as
an integral of the product of the
muon distribution times the magnetic field distribution over the storage
region. Since the moments of the muon distribution couple to the respective 
 multipoles of the magnetic field, either one needs an exceedingly uniform
 magnetic field, or exceptionally good information on the muon orbits in the
 storage ring, to determine $\langle B\rangle_{\mu-dist}$ to sub-ppm
 precision.
  Thus traditional magnetic focusing used in storage rings,
 which involves magnetic quadrupole and higher multipoles, will cause large
 uncertainties in the knowledge of $\langle B\rangle_{\mu-dist}$.  This problem
 was mitigated in the third CERN experiment\cite{blr-CERN3}, and in E821, 
by using electrostatic quadrupoles to provide the vertical focusing, freeing
the magnetic-field design to be as close to a 
uniform dipole field as possible. 
Examination of Eq.~(\ref{eq:blr-omegaa}) shows that for $\gamma = 29.3$, called
``the magic $\gamma$,'' an
electric field will not contribute to $\omega_a$.  The electric-field effect
vanishes for particles with
the central momentum equal to $p_{magic} = 3.09$~Gev/c, and is a small
(sub-ppm) correction for other stored muons~\cite{blr-bennett06}.  

The CERN experiment used a rectangular aperture, which at their
7.3~ppm level of
precision did not cause problems in determining the average field.  However,
the large moments of a rectangular beam were not acceptable for the BNL
experiment, which aimed at a factor-of-twenty improvement.  Thus a circular
beam aperture was chosen for E821, which resulted in a systematic error on 
 $\langle B\rangle_{\mu-dist}$ of 0.03~ppm, certainly adequate for the
 experiment now proposed at Fermilab.  

The experiment consists of repeated fills of the storage ring,  each time
introducing an ensemble of muons into a magnetic storage ring, and then
measuring the two
frequencies $\omega_a$ and $\omega_p$. The muon lifetime
is given by $\gamma \tau_\mu = 64.4~\mu$s, and the data collection period
is typically  $\sim 10$ muon lifetimes in the ring.
The $(g-2)$ precession period is 4.37~$\mu$s, and the 
cyclotron period is 149~ns.
As the $\mu^-$ (or $\mu^+$) decay, $e^-$ ($e^+$) are emitted in the decay
$\mu^- (\mu^+) \rightarrow e^-(e^+) + 
\nu_\mu (\bar \nu_\mu) + \bar \nu_{e} (\nu_e)$.  The high-energy decay
electrons (positrons) carry information on the muon spin direction at the
decay. Thus as the spin turns
relative to the momentum, the number of high-energy decay electrons is
modulated by the frequency $\omega_a$, as shown in
Fig.~\ref{fg:blr-wiggle}.

\begin{center}
\includegraphics[width=8cm]{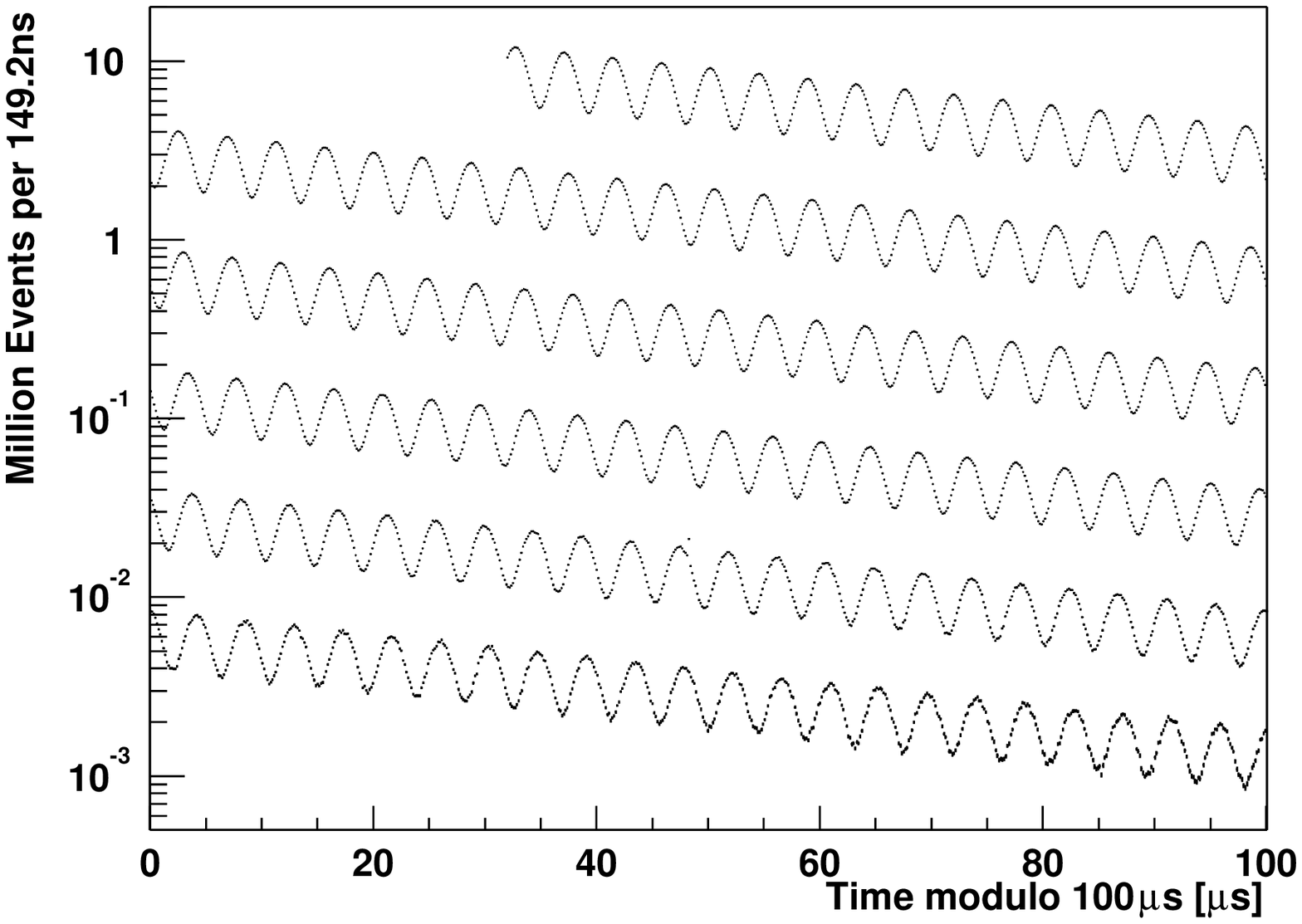}
\figcaption{\label{fg:blr-wiggle} The time spectrum of decay electrons
  from the 2001 E821 running period, when $\mu^-$ were stored. (From
  Ref.~\cite{blr-bennett04} )}
\end{center}

The E821 storage ring was constructed as a ``super-ferric''
magnet~\cite{blr-danby}, meaning that the iron determined the shape of the
magnetic  field. Thus $B_0$ needed to be well below saturation
and was chosen to be 1.45~T.  The resulting ring had
a central orbit radius of 7.112~m, and 12 detector stations were placed
symmetrically around the inner radius of the storage ring.  The 
detector geometry and
number were optimized to detect the high-energy decay electrons, which carry
the largest asymmetry, and thus information on the muon spin direction
at the time of decay. In this design,  many of the
lower-energy electrons miss the detectors, reducing background and pileup.
The electrostatic quadrupoles~\cite{blr-quads} cover 43\%
of the ring, leaving significant gaps for the fast muon
kicker\cite{blr-kickernim}  and other
objects in the ring.

While alternate schemes for measuring $a_\mu$ have been proposed, the magic
$\gamma$ technique has a number of things in its favor. One of the most
important is that since E821, it is quite well understood and offers a
straightforward path to a 0.1~ppm measurement, or perhaps somewhat beyond.
Its features are:
\begin{itemize}
\item high muon polarization and decay asymmetry;
\item large storage ring with ample room for detectors, field mapping,
  kickers, etc.;
\item muon injection, which has been shown to work;
\item rates in the detectors that are easily handled
 with conventional technology;
\item data are fit over many $(g-2)$ cycles, which is a powerful tool to unmask
  systematic errors that depend on time;
\item precision magnetic field techniques  which are well understood;
\item well understood  systematic errors.
\end{itemize}

\section{The Fermilab Proposal: P989}

The Fermilab proposal~\cite{fermilabP989} uses the magic $\gamma$ in
 the  precision storage ring 
developed for E821, with new detectors, electronics,
along with improved magnetic field
measurement and control. Central to the new proposal is
the use of  features unique to Fermilab that will provide copious
 proton bunches of $\sim 10^{12}$ protons at 10 to 20~ms
intervals.  This compares with  $\sim 4-5 \times 10^{12}$ protons
per bunch at BNL, with a maximum of 12 bunches per machine cycle time of 2.7~s.
The effective fill rate at BNL was 4.4~Hz, compared with a projected rate of
18~Hz at Fermilab.  

At BNL, pions 1.7\% above the magic momentum decayed in an 80~m long
FODO line, producing a beam that contained an equal number of pions, muons
and electrons.  A large hadronic ``flash'' accompanied the injection into the
ring causing a significant baseline shift in the detectors near the
injection point.

At Fermilab, the Recycler Ring will be used to re-bunch each proton batch
from the Booster into four bunches with  $\sim 10^{12}$ protons each.  These
will be extracted one at a time to 
 a production target at the location of the present antiproton
target.  The antiproton debuncher ring will be used as
a 900~m long pion decay line. The resulting pion flash will be decreased by a
factor of 20 from the BNL level, and the muon flux will be significantly
increased because of the ability to take zero-degree muons.  The
stored muon-per-proton ratio will be increased by a factor of 5 to 10 over BNL.
Segmented detectors~\cite{blr-McNabb} and new electronics
 should easily be able to handle the
increased data rates per fill of the ring.

The plan is to move the E821 muon storage ring to Fermilab, and install it in a
new building near the existing AP0 hall. The proposal was well received by
the Fermilab Program Advisory Committee, but funding has not yet been
secured.   An optimistic schedule has the ring
moved, re-assembled and shimmed by 2014. We estimate
 that in two years of running on $\mu^+$,
 we could achieve the
goal of the $\pm 0.14$~ppm error. Most of this running would be
simultaneous with NOVA, using the extra Booster batches that cannot
be used by the Main Injector program.  If the Main Injector program is down,
then $(g-2)$ can use the full Booster beam.
  With further running we might be able to
approach the 0.1~ppm level.  During the Project X era, we could achieve a
a comparable error for $\mu^-$.

\section{Summary and Conclusions}

The muon anomalous magnetic moment has played an important role in the
development of the Standard Model, and in constraining theories of physics
beyond the Standard Model.  E821 at the Brookhaven Lab AGS achieved a factor
of 13.5 in precision over the famous CERN experiments of the 1970s, and reached
a relative precision of $\pm 0.54$~ppm. The New Muon $(g-2)$ Collaboration
has proposed to improve the error by a factor of four at Fermilab.
Given the sensitivity of $a_\mu$ to a number of proposed extensions to the
Standard Model, a more precise measurement, especially when combined with
improvements in the knowledge of the hadronic contribution that are on the
horizon, will provide valuable information for the interpretation of new
phenomena that might be discovered at LHC.

\acknowledgments{I wish to thank my many colleagues on E821, and the New
  $(g-2)$ for useful discussions. Thanks to D. Hertzog for comments on
  this manuscript, and to K. Ellis for editorial help.  My participation
in this  work was
  supported in part by the U.S. National Science Foundation.
}

\end{multicols}

\clearpage


\section{References}


\begin{thebibliography}{90}

\vspace{3mm}


\bibitem{blr-Kusch} P. Kusch and H.M Foley, Phys. Rev. {\bf 73}, 250 (1948).


\bibitem{blr-schwinger}  J. Schwinger, Phys. Rev. {\bf 73}, 416L (1948), and
Phys. Rev. {\bf 76} 790 (1949). The former paper contains a misprint
in the expression for $a_e$ that is corrected in the longer paper.

\bibitem{blr-MdRR2007} J.P. Miller, E. de Rafael and B.L. Roberts,
 hep-ph/0703049, and Rep. Prog. Phys. {\bf 70},  (2007) 795-881;

\bibitem{blr-Garwin} R.L. Garwin, D.P. Hutchinson, S. Penman and G. Shapiro,
Phys. Rev. {\bf 118}, 271 (1959).

\bibitem{blr-bennett04} G. Bennett, et al.,  (Muon $(g-2)$ Collaboration),
Phys. Rev. Lett. {\bf 92}, 161802 (2004).

\bibitem{blr-bennett06} G. Bennett, et al.,  (Muon $(g-2)$ Collaboration), 
Phys. Rev. {\bf D73}, 072003 (2006), and references therein.


\bibitem{blr-CODATA08}Peter J. Mohr, Barry N. Taylor and David B. Newell,
Rev. Mod. Phys. {\bf 80}, 633 (2008).

\bibitem{fermilabP989} New Muon $(g-2)$ Collaboration, \hfill \break
http://lss.fnal.gov/archive/test-proposal/0000/fermilab-proposal-0989.shtml


\bibitem{blr-Davier09} M. Davier, A. H\"ocker, B. Malaescu, C.Z. Yuan and
  Z. Zhang, arXiv:0908.4300 Aug. 2009.

\bibitem{blr-PdRV} J. Prades, E. de Rafael and A. Vainshtein, 
in {\it Lepton Dipole Moments}, Advanced Series on Directions in
High Energy Physics -Vol.20, ed. B. Lee Roberts and William J.
 Marciano, World Scientific, p. 303, (2010); and arXiv:0901.0306, Jan. 2009.


\bibitem{blr-Davier09-tau}M. Davier, A. H\"ocker, G. L\'opez Castro,
  B. Malescu, H.X. Mo, G. Toledo S\'anchez, P. Wang, C.Z. Yuan and Z. Zhang,
 arXiv:0906.5443, 2009 and submitted to Eur. Phys. J. 

\bibitem{blr-stockinger}Dominik St\"ockinger, in {\it Lepton Dipole
 Moments}, Advanced Series on Directions in
High Energy Physics -Vol.20, ed. B. Lee Roberts and William J.
 Marciano, World Scientific, p. 393, (2010);
 and J. Phys. {\bf G 34}, R45 (2007).

\bibitem{blr-czmar} Andrzej Czarnecki and William J. Marciano, in  
 {\it Lepton Dipole Moments}, Advanced Series on Directions in
High Energy Physics -Vol.20, ed. B. Lee Roberts and William J.
 Marciano, World Scientific, p. 11, (2010); and Phys. Rev. {\bf D 64}, 
013014 (2001).

\bibitem{blr-whitepaper}David W. Hertzog, et al., arXiv.0705.4617, May 2007.

\bibitem{blr-liu} W. Liu et al., {Phys. Rev. Lett.} {\bf 82}, 711 (1999).

\bibitem{blr-CERN3} J. Bailey, et al.,  Nucl. Phys. {\bf B150}, 1 (1979).

\bibitem{blr-danby} G.T. Danby, et al., Nucl. Inst. and Meth., 
{\bf A 457}, 151-174 (2001).

\bibitem{blr-quads} Y.K. Semertzidis, et al.,  Nucl. Inst. 
and Meth. {\bf A503}, 458-484 (2003).

\bibitem{blr-kickernim}E. Efstathiatis, et al., Nucl.
 Inst. Meth. A {\bf 496}, 8  (2003). 

\bibitem{blr-McNabb} R. McNabb, et al., Nucl. Inst. Meth {\bf A602}, 396
  (2009). 

\end{thebibliography}
\end{document}